\documentclass{nature}
\usepackage[utf8]{inputenc}
\usepackage{graphicx}
\makeatletter
\let\saved@includegraphics\includegraphics
\AtBeginDocument{\let\includegraphics\saved@includegraphics}
\renewenvironment*{figure}{\@float{figure}}{\end@float}
\makeatother
\usepackage{dcolumn}
\usepackage{bm}
\usepackage{color,soul}
\definecolor{purple}{RGB}{127,0, 255}

\usepackage{caption}
\usepackage{ragged2e}
\usepackage{amssymb}
\usepackage{amsmath}
\usepackage{mathtools}
\usepackage{multirow}
\usepackage{lineno}

\title{Observation of Floquet Chern insulators of light} 

\author{Jicheng Jin$^{1,4}$, Li He$^{1,4}$, Jian Lu$^{1,3}$, Lin Chang$^{2}$, Chen Shang$^{2}$, John E. Bowers$^{2}$, Eugene J. Mele$^{1}$, Bo Zhen$^{1}$}

\begin{document}
\maketitle

\begin{affiliations}
 \item Department of Physics and Astronomy, University of Pennsylvania, Philadelphia, Pennsylvania 19104, USA
 \item Department of Electrical and Computer Engineering, University of California, Santa Barbara, California, USA
 \item Current affiliation: Cymer Inc., 17075 Thornmint Ct.
San Diego, California 92127, USA
  \item These authors contributed equally: Jicheng Jin, Li He. 
\end{affiliations}

\section*{Abstract} 
The field of topological photonics studies unique and robust photonic systems that are immune to defects and disorders due to the protection of their underlying topological phases. 
Mostly implemented in static systems, the studied topological phases are often defined in linear photonic band structures.
In this study, we experimentally demonstrate Floquet Chern insulators in periodically driven nonlinear photonic crystals, where the topological phase is controlled by the polarization and the frequency of the driving field. 
Mediated by strong material nonlinearity, our system enters what we call the ``strong Floquet coupling regime", where the photonic Floquet bands cross and open new energy gaps with non-trivial topology as observed in our 
transient sum-frequency generation measurements. 
Our work offers new opportunities to explore the role of classical optical nonlinearity in topological phases and their applications in nonlinear optoelectronics.

\maketitle

\section*{Introduction}

One of the key concepts in the study of light-matter interaction is the strong-coupling regime\cite{thompson_observation_1992,chang_colloquium_2018}, where light and matter degrees of freedom (e.g., atoms\cite{douglas_quantum_2015}, excitons\cite{zhang_photoniccrystal_2018,deng_excitonpolariton_2010a}, phonons\cite{faust_mixing_1966}, plasmons\cite{zhang_surface_2012a}, and magnons\cite{camley_surface_1982}) hybridize and form new quasiparticles known as polaritons, giving rise to a plethora of fascinating phenomena\cite{basov_polaritons_2016}.
Beyond these linear systems in static, a new realm exists in periodically driven (i.e., Floquet\cite{dittrich1998quantum}) nonlinear optical materials, where two orthogonal eigenmodes of the linear system strongly interact, mediated by an external driving field and material nonlinearity.
A characteristic Floquet setup can be a multi-mode optical resonator made of parametric-nonlinear materials periodically driven by a monochromatic field, causing the underlying linear resonances to interact. 
When the nonlinear interaction is strong enough to overcome the losses, the system enters, what we call, the ``strong Floquet coupling" regime, where the linear resonances renormalize into new Floquet eigenmodes, often accompanied by phenomena like normal mode splittings as observed in isolated resonators\cite{xu_floquet_2020a,guo_onchip_2016a,dobrindt_parametric_2008}.  

The theoretical framework of Floquet physics is also applicable to spatially periodic materials and structures, such as crystalline solids and photonic crystals (PhCs), leading to the formation of Floquet-Bloch bands that are bounded in both energy and momentum.
In the strong Floquet coupling regime, new energy gaps (i.e., Floquet gaps) are opened with their topological properties determined by the driving field, paving a new path to engineer band topology\cite{hasan_colloquium_2010a,rudner_band_2020b,peterson_fractional_2020a,hafezi_imaging_2013}.


Floquet topological phases have been studied in various wave systems\cite{rechtsman_photonic_2013,lu_floquet_2021,zhu_timeperiodic_2022}, with most demonstrations in electronics\cite{wang_observation_2013,mciver_lightinduced_2020,zhou_pseudospinselective_2023b,ito2023build}, which exhibit fundamental differences from their photonic counterparts\cite{rechtsman_photonic_2013}.
Whether particle number is conserved or not gives rise to many of these distinctions. 
For example, the conservation of charges in a closed system guarantees a Hermitian eigenvalue problem. 
Accordingly, electronic Floquet topological phases are always concerned with the band topology of real energy spectra. 
In contrast, the number of photons is not necessarily conserved in nonlinear optical processes (e.g., in an optical parametric amplifier, both signal and idler photon number increase due to high-energy pump photons, $\omega_{\rm sig} + \omega_{\rm idl} = \omega_{\rm pump}$) and the corresponding Floquet eigenvalue problems are necessarily non-Hermitian\cite{winn_interband_1999a,he_floquet_2019a}. 
Moreover, even in scenarios with guaranteed real  spectra\cite{he_floquet_2019a} (e.g., when pump photons have low energies, $\omega_{\rm sig} - \omega_{\rm idl} = \omega_{\rm pump}$), photons can be dissipated through the intrinsic loss of Floquet eigenmodes (via material absorption or radiation into free space modes), in contrast to periodically driven electronic systems where interacting electrons will keep absorbing energies from the driving field and dissipating energies to other system degrees of freedom, which eventually leads to an infinite-temperature state where no topological features can be observed\cite{rudner_band_2020b}. 
From an experimental point of view, electronic bands can directly interact with the driving electromagnetic field (e.g., via electric-dipole transitions), while photons rarely interact, except through material nonlinearity that is generally weak.  
Consequently, in practice, one needs PhC modes with long lifetime to observe Floquet topological phases of light.

Here we experimentally demonstrate Floquet Chern insulators of light by driving a nonlinear PhC into the strong Floquet coupling regime and opening Floquet gaps with non-trivial Chern numbers. 
The AlGaAs-on-isolator platform is chosen for its large quadratic nonlinear susceptibilities and low losses in the near-infrared regime, enabling both strong Floquet couplings and long resonance lifetime. 
We start by presenting the Floquet band structures and their corresponding topological properties under driving fields of different polarizations. 
The Floquet band dispersions are experimentally probed in transient sum-frequency generation, 
 featuring non-trivial dependence on the drive.
Fitting these experimental results to a Floquet temporal coupled mode theory we develop, we extract the nonlinear interaction strength between photonic bands and confirm that it indeed overcomes all the losses combined and our system has reached the strong Floquet coupling regime. 
Finally, we experimentally reconstruct our Floquet band structures, showing good agreement with simulation, and demonstrate the emergence of Floquet Chern insulators under circularly polarized drive. 

\section*{Main text}

First, we present the theoretical framework and the design of our optical Floquet Chern insulator. We then show our linear and nonlinear optical measurement results, which include Floquet bands and band-gaps measured under driving fields of various polarizations and frequencies. 

\subsection{Floquet Chern insulators of light identified by symmetry indices}

We begin by introducing the exact configuration of our Floquet Chern insulator, consisting of a nonlinear PhC driven by an external laser (Fig.~1a).  
The nonlinear PhC consists of a square lattice of air cylindrical holes in an Al$_{0.28}$Ga$_{0.72}$As slab on a SiO$_2$ substrate. 
The calculated linear band structure (Fig.~1b) shows the three bands of relevance, including one at higher frequencies and two at lower frequencies forming a quadratic degeneracy at $\Gamma$. 
Along $k_x$ and $k_y$ axis, the two low-frequency bands can be distinguished by their distinctive mirror-symmetry properties: the red (blue) one always radiates $y-$polarized ($x-$polarized) light in the far field. 
See Methods for more details on the material parameters used in the numerical simulation.
When the nonlinear PhC is periodically driven in time by an external laser at an angular frequency $\Omega$, the linear band structure is replaced by Floquet band structures. 
Specifically, due to the quadratic nonlinearity of AlGaAs, the photonic bands produce sidebands (e.g., dashed lines) that shift up and down by multiples of $\Omega$ in the spectrum, similar to sum- and difference-frequency generation processes.  

The polarization and frequency of the driving field determine how these sidebands interact and re-normalize into Floquet bands, as well as their symmetries and topologies. 
For instance, a $x-$polarized drive preserves both $x-$ and $y-$mirror symmetries of the system and can only couple two of the three photonic bands (red and black), but not all three (Fig.~1c).  
This can also be intuitively understood by analyzing the nonlinear susceptibility of AlGaAs, $\chi^{(2)}_{\rm xyz}$. Under a $x$-polarized drive, the $E_y$ component of the red band is coupled to the $E_z$ component of the black band inside the PhC, while neither of them couples to the $E_x$ component of the blue band.
As a result, under a $x-$polarized drive at frequency $\Omega c/2\pi a=0.15$, a Floquet gap is opened along $k_x$ axis, where the red and yellow bands cross, but no gap is opened along $k_y$, where the blue and yellow bands cross. 
The resulting Floquet bands remain connected in the spectrum (i.e., gapless) through two linear (Dirac) degeneracies along $k_y$. 
In contrast, a circularly polarized drive ($\hat{x}+i\hat{y}$) at the same frequency breaks both mirror symmetries and couples all three photonic bands. 
The resulting Floquet band structure features a full gap (shaded in gray in Fig.~1c) between the top and the two bottom Floquet bands, which is topological and characterized by a non-zero Chern number $C=1$. 

The Chern number of the Floquet gap is confirmed in three ways. 
First, it is calculated by integrating the Berry curvature of the Floquet bands over the entire Brillouin zone, following our previous derivation\cite{he_floquet_2019a}. 
More details on the Berry curvature calculations are provided in Section I in the Supplementary Information. 
The Chern number is then confirmed through the bulk-edge correspondence in a super-cell geometry, where a pair of unidirectional Floquet chiral edge states are observed at the edge of the Floquet Chern insulator. 
More details on the FLoquet chiral edge states are provided in Section II in the Supplementary Information. 

In a most straightforward way, the non-trivial topology is also confirmed using the symmetry indices at high-symmetry $k-$points. 
Specifically, the driving field, regardless of its polarization, always preserves the space-time compound symmetry\cite{jin_floquet_2022} of $\Tilde{C}_2=C_2^z\hat{T}_{T/2}$. 
Here, $C_2^z$ is the 180-degree spatial rotation around the $z-$axis, and $\hat{T}_{T/2}$ is the temporal translation by half of the modulation period ($T = 2\pi / \Omega$). 
Four points in the $k-$space ($\Gamma$, $X$, $Y$, $M$) are invariant under $\Tilde{C}_2$, and their $\Tilde{C}_2$ index must be $\pm1$ since $\Tilde{C}_2^2=1$. 
The Chern number of the highest-frequency Floquet band can be related to its $\Tilde{C}_2$ indices at these four points: 
\begin{equation}
    e^{i \pi C} = \Tilde{C}_2 (\Gamma) \Tilde{C}_2 (X) \Tilde{C}_2 (Y) \Tilde{C}_2 (M). 
\end{equation}
Under a linearly polarized (e.g., $x-$) drive, the Floquet bands always remain gapless, preventing one from defining a Chern number using Eq.~1. 
In contrast, the Floquet gap is fully opened under a circular drive, and the gap Chern number can be further simplified as $ e^{i \pi C} = \Tilde{C}_2 (\Gamma) \Tilde{C}_2 (M) = -1$, confirming that the Floquet gap is indeed topologically non-trivial ($C\neq 0$).  
This is consistent with the fact that time-reversal symmetry is preserved under a linear polarized drive but broken under an elliptically (e.g., circularly) polarized drive in AlGaAs\cite{jin_floquet_2022}. 
Section III in the Supplementary Information provides more details on the symmetry indices and how topological phase transition can happen in the polarization space of the driving field in this particular system.   

\subsection{Linear band characterizations of AlGaAs PhCs} 
The AlGaAs PhCs (Fig.~2a) are fabricated following the procedures of wafer bonding, electron-beam lithography, and reactive ion etching (see Methods for more details).  
The linear band structure is measured through angle- and polarization-resolved linear reflection spectroscopy (Fig.~2b), which shows the three photonic bands of interest and their corresponding far-field polarizations ($x$ or $y$) along the $k_x$ and $k_y$ axis. 
By fitting the reflection data to a temporal coupled mode theory model\cite{wonjoosuh_temporal_2004}, the linear band structure is reconstructed (Fig.~2c), including both the resonance frequencies ${\rm Re}(\omega)$ and the decay rates ${\rm Im}(\omega)$.  
Section IV in the Supplementary Information provides more details on the linear temporal coupled mode theory and the fitting procedures.
 
\subsection{Observation of nonlinear Floquet bands} 

To characterize the Floquet band structure, a pump-probe setup is developed (Fig.~3a). 
The nonlinear PhC is periodically driven by a narrow-band pump field at wavelength $\lambda_{\rm P}$ (pulse duration $\sim$1.8 picoseconds) and incidents from the normal direction.
Meanwhile, the PhC resonances are excited by a broadband probe beam ($\sim$169 femtoseconds) centered at wavelength $\lambda_{\rm B}$. 
Both the pump and probe lasers are seeded from the same oscillator and amplifier (Pharos, Light Conversion) to minimize relative time jitter. 
The pump and probe beams are properly delayed in time to achieve optimal temporal overlap on the sample.
The pump beam, probe beam, and quadratic nonlinearity of AlGaAs work together to populate the Floquet bands, whose far-field radiation is captured by the same lens and recorded with a monochromator and a 2D CCD. 
This setup can measure both transient sum-frequency generation (SFG) at wavelength $\lambda_{\rm S}$ and transient reflection signals at $\lambda_{\rm B}$, separable via a dichroic mirror. 
Section V in the Supplementary Information provides more details on the nonlinear characterization setup. 

Examples of our measured Floquet bands are shown in Fig.~3b.
Here, the mid-infrared pump field ($\lambda_{\rm P} \approx 3\mu$m) is $x-$polarized. The near-infrared probe field ($\lambda_{\rm B} \approx 1\mu$m) is $y-$polarized, exciting the red band in Fig.~1b and 2c. 
Strong SFG emission is observed from the sidebands in the visible regime ($\lambda_{\rm S}\approx~$760 nm). 
Meanwhile, the transient SFG spectrum is strongly modulated by the linear photonic resonances (dashed blue line), corresponding to the black band in Fig.~1b and Fig.~2c. 
As expected, the sideband frequency (color map) keeps increasing with the pump frequency --- as $\lambda_{\rm P}$ decreases from 3.2$\mu$m to 3.02$\mu$m --- while the frequency of the linear resonance (dashed blue line) remains largely unchanged.

\subsection{Reaching the strong Floquet coupling regime} 
We experimentally determine the nonlinear coupling strength between photonic bands and confirm that our system has reached the strong Floquet coupling regime, opening Floquet gaps in the spectrum. 
To quantify the coupling strength $g$, we analyze the intensity dependence of the transient SFG on pump power (Fig.~4a). 
In these measurements, the pump wavelength is fixed at $\lambda_{\rm P} = 3.05\mu$m, and we record the intensity of the transient SFG at $760$nm, where the two Floquet bands cross (highlighted by a white box in Fig.~3b), as the pump power increases from 0 to 45mW. 
As shown, the transient SFG intensity, $I_{\rm SFG}$, reaches its maximum value of $I_{\rm SFG}^{\rm max}$ when the pump power is at $P=21$mW and is reduced at lower or higher pump. 
Similar power dependence has been observed in previous literature on resonance-based SFG\cite{guo_onchip_2016a} and is also captured by our Floquet temporal coupled mode theory.
Specifically,  $I_{\rm SFG}$  can be expressed as:
\begin{equation}
\frac{I_{\rm SFG}}{I_{\rm SFG}^{\rm max}} = \frac{4C}{(1+C)^2}, 
\end{equation}
where $C$ is the cooperativity of the system, defined as $C = g^2/(\gamma_1 \gamma_2)$. 
Meanwhile, $\gamma_{1,2} = {\rm Im}(\omega_{1,2}) $, which equals half the resonance linewidth, represents the loss of the two involved resonances, respectively.
Following Eq.~2, $I_{\rm SFG}$ reaches its maximum under the condition of $C=1$, which is satisfied at $P=21$mW. 

Both coupling strength $g$ and cooperativity $C$ increase with pump power: $g \propto \sqrt{P}$ and $C \propto P$. 
Our largest experimentally achieved coupling strength is $g_{\rm max}/2\pi = 185$GHz and the highest cooperativity is $C_{\rm max} = 2.06$, both under the pump power of $P=45$mW (Fig.~4b). 
This nonlinear coupling strength is larger than all losses combined in our system, $g_{\rm max}/2\pi>\sqrt{(\gamma_1^2 + \gamma_2^2)/2}/2\pi = 129.1$GHz, which proves that our system has entered the strong Floquet coupling regime, even by the most stringent criteria used in the literature of cavity quantum electrodynamics \cite{peng_what_2014a,reithmaier_strong_2004,rudner_band_2020b,osti_121765} 
Section VI and VII in the Supplementary Information provides more details on the Floquet temporal coupled mode theory model and the fitting procedure. 

\subsection{Observation of nonlinear Floquet Chern insulators} 
The complete Floquet band structure can be reconstructed 
by fitting the experimental data (linear reflection and transient SFG spectra) to our Floquet temporal coupled mode theory and extracting the nonlinear coupling strength over a wide range of momenta.  
As a result, the reconstructed Floquet band structure under a $x-$polarized driving field is shown in Fig.~5a, which features a Floquet gap along the $k_x$ axis as well as a pair of linear (Dirac) degeneracies along the $k_y$ axis --- both features are in good agreement with our theoretical prediction in Fig.~1b and the experimental results in Fig.~4b.  
In contrast, the reconstructed Floquet band structure under a circularly polarized drive is fully gapped along both directions (Fig.~5b). 
The resulting  system is a Floquet Chern insulator, corresponding to our calculations in Fig.~1c. 
Section IX in the Supplementary Information provides more details on the reconstruction of the Floquet band structure from our Floquet temporal coupled mode theory. 

\section*{Discussion and conclusion}

Our pump-probe setup is designed to maximize Floquet physics arising from quadratic nonlinearity while minimizing the impact of unwanted nonlinear processes. 
For example, our pump pulse duration is chosen to be about 1.8 picoseconds, which is too short for thermal nonlinearities to take effect. 
Meanwhile, this time scale is comparable to the lifetime of the photonic resonances involved, ensuring strong interactions between the pump field and the resonances. 
Our pump field (\(\sim \)0.3GV/m) guarantees quadratic nonlinearities to dominate over other (e.g., cubic) nonlinearities in our system.
Finally, our SiO$_2$ substrate is chosen to be JGS3 to avoid MIR pump absorption and cubic nonlinearity. 
 
In current our experimental results, the largest achieved nonlinear cooperativity $C_{\rm max}\approx2.1$ is limited by two factors. 
First, our peak pump field is close to the damage threshold of AlGaAs at this wavelength for this pulse duration. 
To further improve the damage threshold, an even lower-frequency driving field (possibly in the far-infrared regime) is needed. 
On the other hand, the material absorption in AlGaAs beyond its direct bandgap at longer wavelengths limits the lifetime of the resonances, which is consistent with the literature.\cite{michael_wavelength_2007a}.
Changing to another material with lower absorption losses but comparable nonlinearities and damage thresholds can be beneficial in further improving the resonance lifetime and, therefore, nonlinear cooperativity.

In summary, we report the experimental observation of Floquet Chern insulators in driven nonlinear photonic crystals, where both the Floquet band structures and their topological properties can be controlled by the driving field. 
Our system enters the strong Floquet coupling regime, as the nonlinear coupling strength, mediated by the strong pump field and material nonlinearity, exceeds all loss mechanisms combined. 
The Floquet band structures are experimentally reconstructed from our linear and nonlinear experimental results, featuring Floquet energy gaps due to the pump and showing good agreement with simulation results.    
Our work paves the way for future explorations of Floquet topological phases in driven nonlinear systems and opens new possibilities in implementing on-chip topologically robust photonic devices and energy-efficient non-reciprocal devices. 

\section*{Methods}

\subsection{Numerical simulation details.} Without a driving field, the PhC band structure is calculated using a finite element method similar to our previous calculations\cite{he_floquet_2019a}. 
The refractive index of Al$_{0.28}$Ga$_{0.72}$As is set to be $n=3.47$ and 3.3 at 760nm and 1$\mu$m, respectively. The refractive index of SiO$_2$ substrate is set to be $n=1.46$. For Floquet band calculations, the quadratic nonlinearity of Al$_{0.28}$Ga$_{0.72}$As is $\chi^{(2)}_{xyz} = \chi^{(2)}_{yzx} = \chi^{(2)}_{zxy} = 200$pm/V, which is consistent with previous literature\cite{chang_strong_2019a,chang_ultraefficient_2020,zhu_integrated_2021a}

\subsection{Sample fabrication.} 
An AlGaAs wafer with an 85 nm thick Al$_{0.28}$Ga$_{0.72}$As film and a 500nm thick Al$_{0.8}$Ga$_{0.2}$As sacrificial layer is grown on a 500$\mu$m thick GaAs substrate. 
The thin film side is flip-bonded onto a SiO$_2$ substrate, which is chosen to be JGS3 to minimize MIR absorption and cubic nonlinearity.  
The GaAs substrate is then removed through mechanical polishing and wet etching with H$_2$O$_2$:NH$_4$OH (30:1). 
We then remove the Al$_{0.8}$Ga$_{0.2}$As sacrificial layer  with dilute hydrofluoric $2.5\%$ acid and deposit a 5nm-thick Al$_2$O$_3$ layer through ALD for passivation. 
For lithography, the chips are spin-coated with a 300nm-thick ZEP 520A and the patterns are defined with Raith EBPG 5200, with a cold development at -6$^\circ$ C in o-xylene. 
The pattern on the photoresist is then transferred onto the AlGaAs layer by ICP  etching (Oxford Corba) with a mixture of BCl$_3$/Cl$_2$/Ar. 
In the end, the sample is passivated with a 5nm-thick layer of Al$_2$O$_3$ by ALD.

\subsection{Optical measurement setups.} All presented experimental results are from angle-resolved and polarization-resolved spectroscopy measurements. 
The sample is mounted onto a JGS3 substrate with anti-reflection coating with optical gel to eliminate spectral fringes due to multiple reflections.
The entire sample is then placed at the focal plane of an achromatic lens with a focal length of 4cm.  
The confocal setup converts the momentum space of the PhC into real space, which is then mapped onto the entrance slit of the monochromator using relay optics. Angle- and polarization-resolved spectrum along one momentum direction is directly read out with a 2D CCD. 
By scanning the position of the entrance slit, measurements along both momentum directions can be performed.
The etaloning effect of the CCD is not significant around 760nm but is clearly visible near 1$\mu$m. It is calibrated out following standard procedures \cite{sanz-arranz_amorphous_2017,massie_calibration_2022}.
Different light sources are used for different measurements. For linear band characterization, a femtosecond optical parametric amplifier is used. For nonlinear measurements, a picosecond MIR optical parametric amplifier is used to drive the sample, while a NIR femtosecond probe laser is used to excite the resonances.




\subsection{Data availability} The data within this paper are available from the corresponding author upon request. 

\subsection{Acknowledgments} This work was partly supported by the U.S. Office of Naval Research (ONR) through grant N00014-21-1-2703, the Army Research Office under award contract W911NF-19-1-0087, and DARPA under agreement HR00112220013. 
Work by E.J.M is supported by the Department of Energy under grant DE-FG02-84ER45118.

\subsection{Author Contributions} 
J.J., L.H., J.L., and B.Z. conceived the project. J.J. and L.H. performed numerical simulations. J.J. fabricated the sample assisted by L.H. L.C. C.S. and J.B. provided bonded wafers. J.J., L.H. and J.L. performed the experiment. J.J., L.H., and B.Z. wrote the paper with input from all authors. All authors discussed the results. B.Z. supervised the project. 

\subsection{Competing interests} The authors declare no competing interest.

\subsection{Correspondence} Correspondence should be addressed to B.Z. (email: bozhen@sas.upenn.edu).

\clearpage

\begin{figure}[ht]%
\centering
\includegraphics[width=\textwidth]{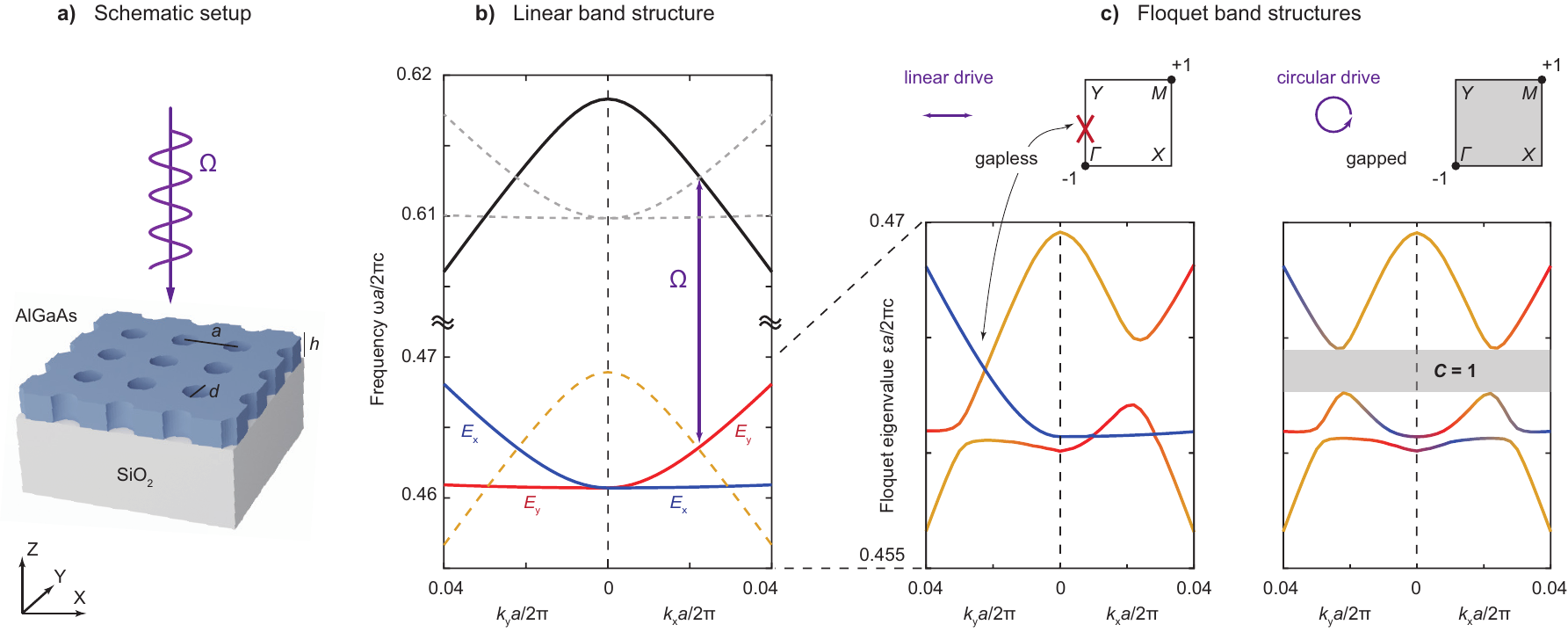}
\caption*{\textbf{Fig.~1 $|$ Floquet Chern insulators of light identified by symmetry indices.}
\textbf{a.} Schematic of a nonlinear (AlGaAs) photonic crystal (PhC) periodically driven by an external laser at frequency $\Omega$. 
The geometric parameters are $h/a = 0.18$ and $d/a = 0.2$. 
\textbf{b.} The linear band structure features three bands of relevance: one at higher frequencies (black line) and the other two at lower frequencies (red and blue).
\textbf{c.} A $x-$polarized drive couples the photonic bands via the quadratic nonlinearity of AlGaAs along the $k_x$ axis, but not along the $k_y$ axis, leading to a pair of linear (Dirac) degeneracies.
\textbf{d.} A circularly polarized drive opens a full Floquet gap (shaded in gray), resulting in a Floquet Chern insulator of light.
The topological nature of the highest-frequency Floquet band is confirmed by its $\Tilde{C}_2$ symmetry indices: $-1$ at $\Gamma$ and $+1$ at $M$.
}\label{Fig1}
\end{figure}

\begin{figure}[ht]%
\centering
\includegraphics[width=\textwidth]{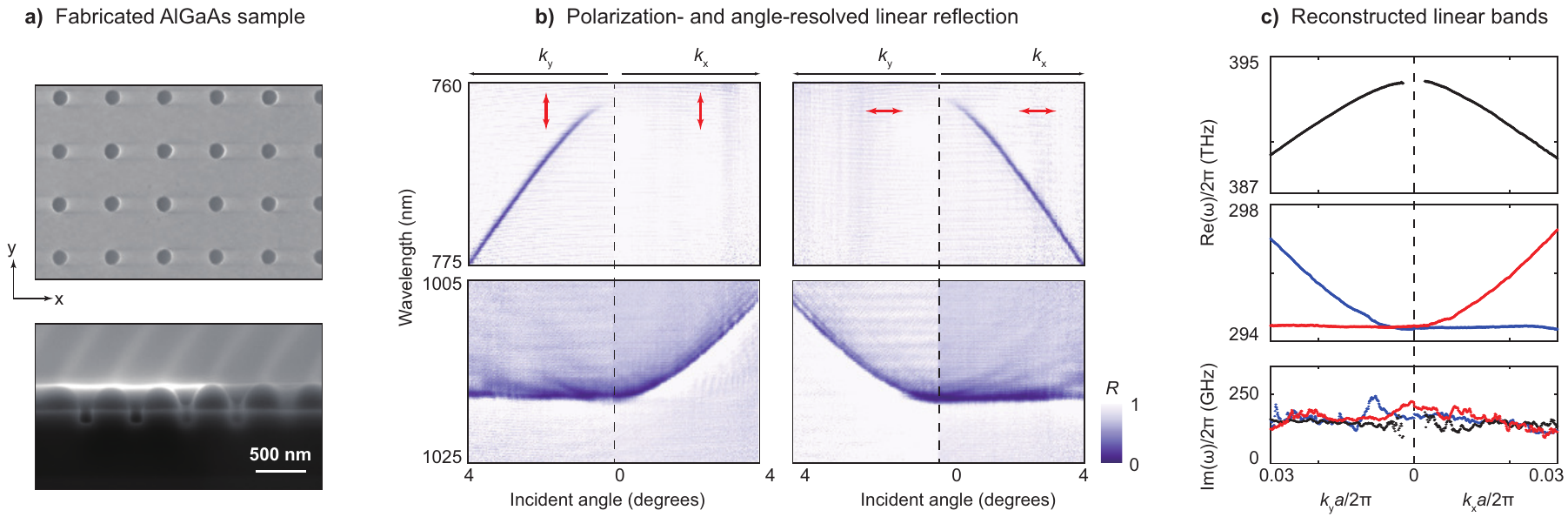}
\caption*{\textbf{Fig.~2 $|$ Linear characterization of AlGaAs PhCs.}
\textbf{a.} SEM images of fabricated AlGaAs PhCs with $a=465$ nm, $d=100$ nm, and $h=85$ nm. 
\textbf{b.} Angle-resolved linear reflection measurements along the $k_x$ and $k_y$ axis under vertically and horizontally  polarized excitations, featuring the three bands of interest. 
\textbf{c.} Experimentally extracted resonance frequencies and decay rates of the three photonic bands from linear reflection measurements.
}\label{Fig2}
\end{figure}

\begin{figure}[ht]%
\centering
\includegraphics[width=\textwidth]{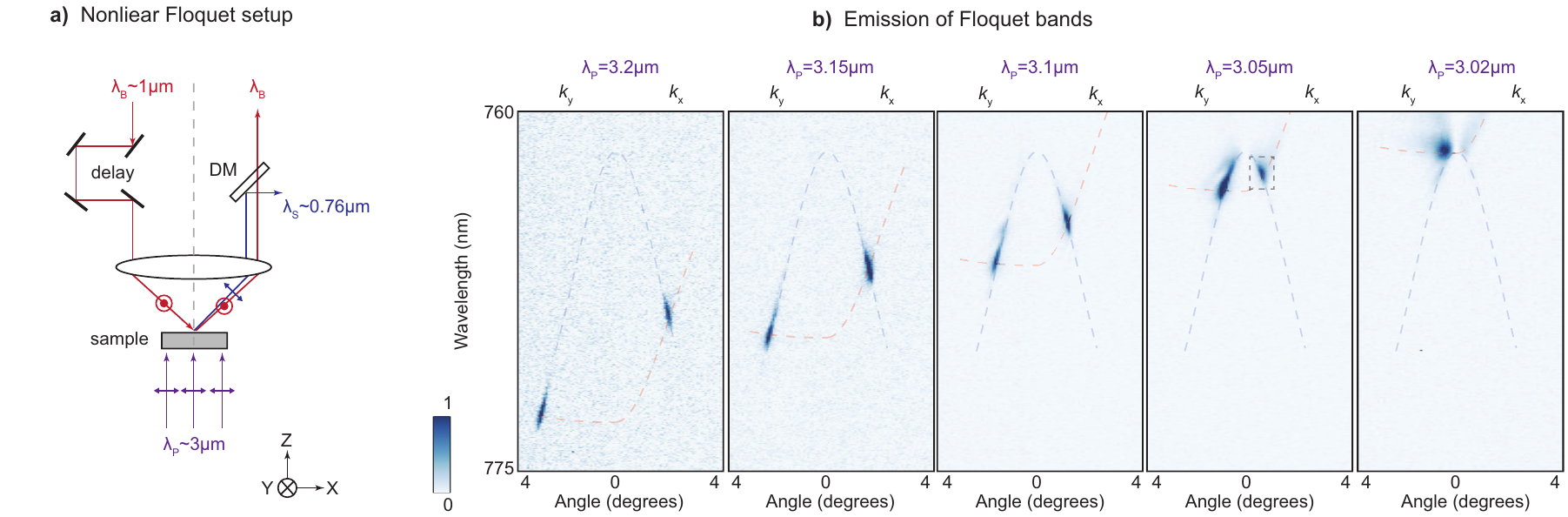}
\caption*{\textbf{Fig.~3 $|$ Observation of nonlinear Floquet bands.} 
\textbf{a}, Schematics of the nonlinear Floquet measurement setup. 
A $x-$polarized pump laser (purple) with a wavelength $\lambda_{\rm P}$ periodically drives the nonlinear PhC, while the probe beam ($\lambda_{\rm B}$, red) excites the linear resonances.
Angle-resolved transient sum-frequency generation ($\lambda_{\rm S}$, blue) is recorded using a monochromator and a 2D CCD. 
\textbf{b}, 
The transient sum-frequency generation spectra clearly show the dispersion of the Floquet bands (dashed lines). 
Their frequencies increase with the pump frequency, as $\lambda_{\rm P}$ changes from $3.2\mu$m to $3.02\mu$m.  
Meanwhile, the linear band remains fixed in frequency (dashed blue line). 
DM, dichroic mirror. 
}\label{Fig3}
\end{figure}

\begin{figure}[ht]%
\centering
\includegraphics[width=0.5\textwidth]{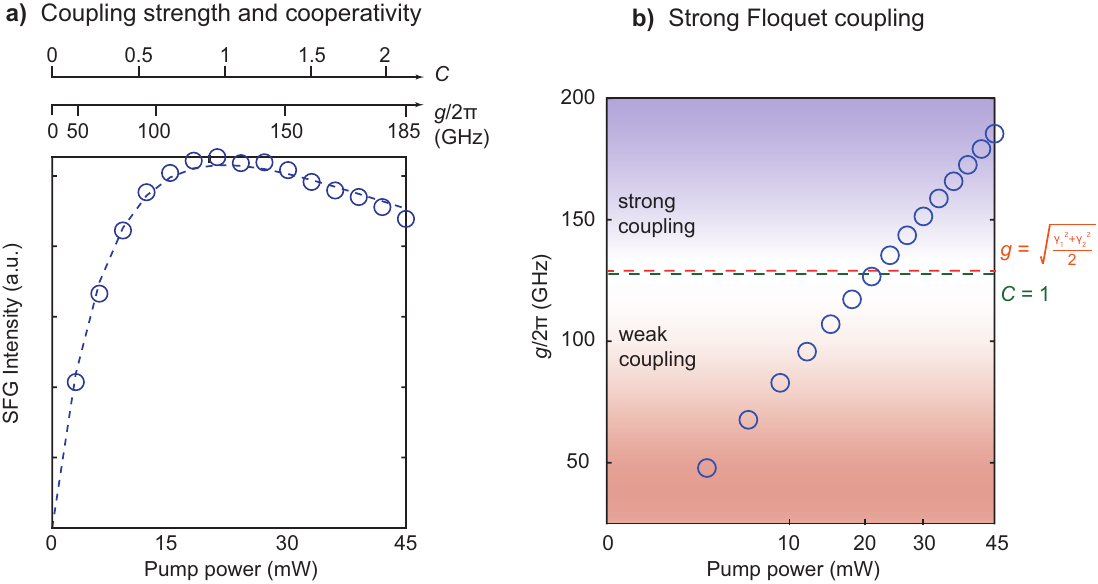}
\caption*{\textbf{Fig.~4 $|$ 
Reaching the strong Floquet coupling regime.} 
\textbf{a}, 
Both nonlinear coupling strength between photonic resonances ($g$) and cooperativity ($C$) are extracted from measured transient SFG intensity $I_{\rm SFG}$ under different pump powers  (circles). 
\textbf{b}, 
The nonlinear coupling strength $g$ increases with the pump power and reaches its maximum value of $g_{\rm max}/2\pi = 185 $GHz under the pump power of $P=45$mW. This nonlinear coupling strength exceeds all loss mechanisms combined (different definitions are shown in dashed lines), proving that our system enters the strong Floquet coupling regime and new Floquet gaps are opened.  

}\label{Fig4}
\end{figure}

\begin{figure}[ht]%
\centering
\includegraphics[width=0.5\textwidth]{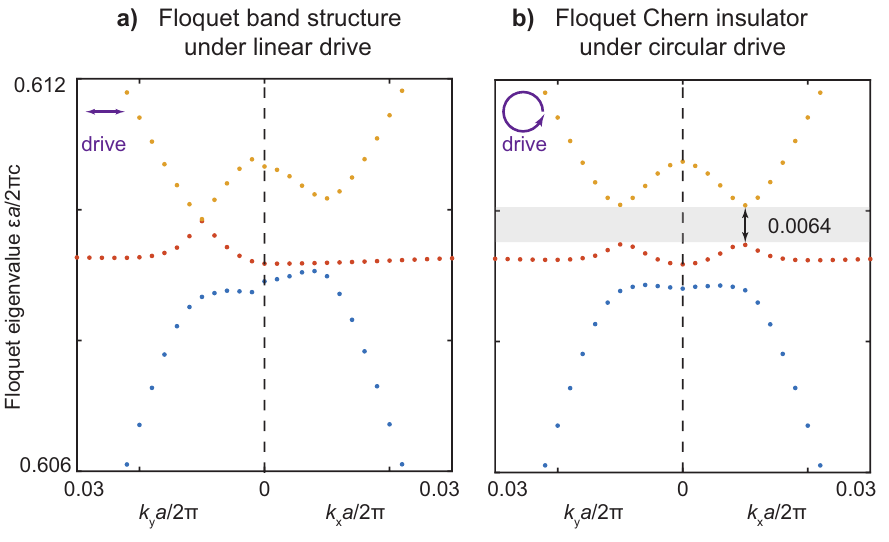}
\caption*{\textbf{Fig.~5 $|$ Observation of Floquet Chern insulators of light.} 
\textbf{a.} 
The Floquet band structures are experimentally reconstructed from linear and nonlinear spectra.
Under a $x-$polarized drive, the Floquet band structure features a Floquet gap along $k_x$ and a Dirac point along $k_y$. 
\textbf{b.} In contrast, under a circularly polarized drive, the Floquet gap is opened along both axes (shaded in gray), corresponding to the nonlinear Floquet Chern insulator in Fig.~1c. 
}\label{Fig5}
\end{figure}



\clearpage

\section*{References}
\bibliographystyle{naturemag}
\bibliography{reference}

\end{document}


\maketitle

\begin{affiliations}
 \item Department of Physics and Astronomy, University of Pennsylvania, Philadelphia, Pennsylvania 19104, USA
 \item Department of Electrical and Computer Engineering, University of California, Santa Barbara, California, USA
 \item Current affiliation: Cymer Inc., 17075 Thornmint Ct.
San Diego, California 92127, USA
 \item Current affiliation: State Key Laboratory of Advanced Optical Communications System and Networks, School of Electronics, Peking University, Beijing, China.
  \item These authors contributed equally: Jicheng Jin, Li He. 
\end{affiliations}

\section{Berry Curvature}

\section{Chiral Edge States}

\section{Symmetry indices}

\section{linear TCMT}

\section{Floquet TCMT}

\section{Fitting and ODE}

\section{Experimental Setup}

\section{reconstruction of Floquet band}

\section*{References}
\bibliographystyle{naturemag}
\bibliography{reference}